\documentstyle[a4,12pt,epsf]{article}
\begin{document}
\pagestyle{empty} 
\hoffset 0.25in
\setcounter{page}{1}
\begin{flushright}
CERN-TH/95-59 \\  ROME prep. 94/1056 \\ LPTHE Orsay-94/88 \\  SHEP prep 95/11
\end{flushright}
\newcommand{\be}{\begin{equation}}
\newcommand{\ee}{\end{equation}}
\newcommand{\bea}{\begin{eqnarray}}
\newcommand{\eea}{\end{eqnarray}}
\newcommand{\nn}{\nonumber}
\newcommand{\muh}{\hat\mu}
\newcommand{\dlr}{\stackrel{\leftrightarrow}{D} _\mu}
\newcommand{\vnew}{$V^{\rm{NEW}}$}
\newcommand{\vecp}{$\vec p$}
\newcommand{\dof}{{\rm d.o.f.}}
\newcommand{\prd}{Phys.Rev. \underline}
\newcommand{\pl}{Phys.Lett. \underline}
\newcommand{\prl}{Phys.Rev.Lett. \underline}
\newcommand{\np}{Nucl.Phys. \underline}
\newcommand{\vvp}{v_B\cdot v_D}
\newcommand{\dl}{\stackrel{\leftarrow}{D}}
\newcommand{\dr}{\stackrel{\rightarrow}{D}}
\newcommand{\mev}{{\rm MeV}}
\newcommand{\gev}{{\rm GeV}}
\newcommand{\calp}{{\cal P}}
\pagestyle{empty}
\begin{center}
{\LARGE
{\bf A lattice study of the  exclusive \\ 
$B\rightarrow K^* \gamma$ decay amplitude,
\\  using  the Clover action  at $\beta=6.0$
}}
\end{center}
\vskip 0.2cm
\centerline{\bf{
As. Abada$^{\rm a}$,  
Ph. Boucaud$^{\rm a}$, N. Cabibbo$^{\rm b}$, M. Crisafulli$^{\rm b}$,
J.P. Leroy$^{\rm a}$,}}
\centerline{\bf{
 V. Lubicz$^{\rm c}$, G. Martinelli$^{\rm d,1}$,
F. Rapuano,$^{\rm b}$, M. Serone$^{\rm e}$, N. Stella$^{\rm f}$}}
\centerline{and}
\centerline{\bf{
A. Bartoloni$^{\rm b}$, C. Battista$^{\rm b}$, 
S. Cabasino$^{\rm b}$, E. Panizzi$^{\rm b}$, P.S. Paolucci$^{\rm b}$, }}
\centerline{\bf{   
R. Sarno$^{\rm b}$, G.M. Todesco$^{\rm b}$, 
M. Torelli$^{\rm b}$, P. Vicini $^{\rm b}$.}}
\vskip 0.15cm
\centerline{The APE Collaboration}
\vskip 0.15cm       
\centerline{$^{\rm a}$ LPTHE, Univ. de Paris XI, 91405 Orsay, France$^2$.}
\centerline{$^{\rm b}$ Dip. di Fisica, Univ. di Roma \lq La Sapienza\rq,}
\centerline{ and INFN, Sezione di Roma, P.le A. Moro, 00185 Rome, Italy.}
\centerline{$^{\rm c}$ Dept. of Physics, Boston University, Boston MA 02215,
USA.}
\centerline{$^{\rm d}$ Theory Division, CERN, 1211 Geneva 23, Switzerland.}
\centerline{$^{\rm e}$ SISSA-ISAS, Via Beirut 2, 34014 Trieste and}
\centerline{Sezione INFN di Trieste, Via Valerio 2, 34100 Trieste, Italy.}
\centerline{$^{\rm f}$ Dept. of Physics, University of Southampton,}
\centerline{Southampton SO17 1BJ, UK.}

%*******************************************************************
\begin{abstract}
We present the results of a numerical calculation of the  $B\rightarrow
K^* \gamma$ form factors. The results have been obtained by studying
the relevant 
correlation functions at $\beta=6.0$, on an $18^3 \times 64$ lattice,
using the ${\rm O(a)}$-improved fermion action, in the quenched
approximation. From the study of the matrix element $ \langle K^*\vert \bar s
\sigma_{\mu\nu} b \vert B\rangle$ we have obtained
the form factor   $T_1(0)$ which controls the exclusive decay rate.
 The results are compared with the recent results from CLEO.
We also discuss the compatibility between 
 the scaling laws predicted by the Heavy
Quark Effective Theory (HQET) and  pole dominance, by studying
the mass- and $q^2$-dependence of the form factors. 
From our analysis, it appears that  the form factors
follow a mass behaviour compatible with the predictions of the HQET  and that
the $q^2$-dependence of $T_2$ is weaker than would be predicted 
by  pole dominance. 
\end{abstract}
%*******************************************************************
\vskip 0.15 cm
\begin{flushleft} 
$^1$ On leave of absence  from Dip. di Fisica, Universit\`a
degli Studi ``La Sapienza", Rome, Italy. \\
$^2$ Laboratoire associ\'e au 
 Centre National de la Recherche Scientifique.
\end{flushleft} 
\vfill\eject
\pagestyle{empty}\clearpage
\setcounter{page}{1}
\pagestyle{plain}
\newpage
\pagestyle{plain} \setcounter{page}{1}
 
%*******************************************************************

\section{Introduction}
\label{sec:intro}
An important class of  $B$-meson decays is given by 
the  weak radiative decays $B\rightarrow
X_s\gamma$, where $X_s$ is a strange hadronic state and where the
emitted photon is real. These decays are extensively studied  because they  
provide, through loop effects, interesting information on
the Standard Model parameters (e.g.  the combination of CKM matrix elements 
$|V_{ts}V_{tb}^*|$).  
Short-distance physics puts in the foreground an effective
magnetic interaction, $b \rightarrow s\gamma$, originating from 
the so-called penguin diagrams.
In these diagrams, the top quark dominates, whereas the
charm and up quark contributions are  suppressed by powers of the quark masses. 

Radiative decays are also sensitive to physics beyond the Standard
Model, through possible extra particles (SUSY particles,
extra charged Higgs bosons, ...) contributing to the  loops. 
Should there be deviations between
the expected decay rates and the measured values, these could be 
a manifestation of new physics.

Among  radiative decays, the process $B\rightarrow K^* \gamma$
has received  an increasing attention, because of the experimental
measurement  of the $B\rightarrow K^* \gamma$ branching ratio, 
performed by the CLEO collaboration \cite{cleo1}: $BR(B\rightarrow K^*
\gamma)=(4.5\pm1.5\pm0.9)\times 10^{-5}$. More recently, the same 
collaboration has also measured  the inclusive rate, obtaining
$BR(B\rightarrow X_s\gamma)=(2.32\pm 0.51\pm 0.29 \pm 0.32)\times 10^{-4}$ \cite{cleo2}.

Several  methods  have been employed
to predict the inclusive $B\rightarrow X_s \gamma$  and exclusive 
  $B\rightarrow K^*\gamma$ decay rates: 
Heavy Quark Effective Theory (HQET)\cite{hqet}, QCD
sum rules
\cite{Dominguez}--\cite{ali},
and quark models \cite{qm1}--\cite{qm3}. For the exclusive decay,
 the theoretical uncertainty,
which was originally of more than two orders of magnitude, has been greatly reduced
by the more recent studies. Still, there is a large
spread between the different results.  Lattice QCD
offers the possibility to investigate rare $B$ decays from first principles. 
The feasibility of the lattice approach was first demonstrated  in 1991,
by the work of Bernard, Hsieh and Soni \cite{hsieh}.
Further results have been obtained  by the UKQCD collaboration \cite{ukqcd} and
by the LNAL collaboration \cite{gupta}. 

In this paper, we present a lattice calculation of the  form factors
 $T_1$ and $T_2$ relevant for  exclusive $B \to K^* \gamma$ decays, 
and of the dependence of these form factors on the heavy-quark  mass and on
 the squared momentum 
transfer $q^2$. 
Preliminary results of our study can be found
in ref. \cite{abadaglasgow}. 
The study of the dependence of the form factors on the
heavy-quark mass  provides a good test of the validity of the
scaling laws predicted by HQET, in
a region of masses around the charm  quark mass.
Particular attention has been devoted to the understanding of these 
scaling laws and their relations to the $q^2$-dependence.
This point will be discussed in detail in section \ref{sec:q2d}.

In the same context, there is a very interesting challenge:
assuming HQET and $SU(3)$ symmetry, the hadronic matrix elements $\langle
K^*(\eta,k)|\bar s \sigma^{\mu\nu}q_\nu b |B(p)\rangle$ are related to
those relevant for  $B$-meson semileptonic decays:
$\langle \rho(\eta,k)|\bar s \gamma^{\mu} b |B(p)\rangle$ and
$\langle \rho(\eta,k)|\bar s \gamma^{\mu}\gamma_5 b |B(p)\rangle$
\cite{wise}--\cite{alain}. This implies relationships
between the semileptonic  and radiative form factors.
These relations can be proved, in the infinite mass limit,
near the zero recoil point, i.e. $q^2 \sim q^2_{max}$.
 A complete test of these relationships on the lattice is
 a very interesting check of QCD dynamics.

With some extra assumptions  one can extend the HQET relations, to the 
region of small  $q^2$  \cite{alain}--\cite{colangelo2}. 
As an example,  consider the ratio
\be
{R(B\rightarrow K^*\gamma)\over {d\Gamma(B\rightarrow \rho \ell
\bar\nu_\ell)\over dq^2}\vert _{q^2=0}}={192\pi^3\over G_F^2} {1\over
|V_{ub}|^2}{(m_B^2-m_{K^*}^2)^3\over
(m_B^2-m_{\rho}^2)^3}{m_b^3\over(m_b^2-m_s^2)^3}|{\cal I}|^2\quad , \label
{Relation}\ee 
where \be R(B\rightarrow K^*\gamma)={\Gamma(B\rightarrow
K^* \gamma)\over \Gamma(B\rightarrow X_s \gamma)}\quad , \quad
{\cal I}={(m_B+m_\rho)\over (m_B-m_{K^*})}{2 \, T_1(0)\over
A_0^{B\rightarrow\rho}(0)}\label {t1eta0}\ee
and $A_0^{B\rightarrow\rho}(0)$ is
one of the six semileptonic form factors (see for example
\cite{lubicz}).  The quantity $\cal I$ can be measured  on the
lattice. According to \cite{alain}--\cite{colangelo2}, 
${\cal I}$  should be close to 1.
Unfortunately, the quality of our numerical results
 does not allow  a calculation 
of ${\cal I}$ at present. 
For this reason,  we only focus here on $B\rightarrow K^*\gamma$
decay.

\par We present  results obtained by assuming different scaling laws, in
the heavy-meson mass, for the form factors at $q^2=0$. These scaling laws
will be  discussed in detail below. The two possibilities correspond
to a pole-dominance behaviour in $q^2$
either for $T_1$ or for  $T_2$ and lead to quite
different results.

If we assume that $T_1$ follows the pole-dominance behaviour, we expect a
scaling law of the form $T_1(q^2=0)=T_2(q^2=0) \sim m^{-1/2}$, where $m$
 is the mass of the heavy quark.
In this case, following ref. \cite{ciubsg}, we obtain 
\bea T_1(q^2=0)&=& 0.23 \pm 0.02 \pm 0.02  \, , \nn \\
R &=& 0.31 \pm 0.12     \, , \nn \\
BR(B \to K^* \gamma)&=& (7.4 \pm 1.4^{+2.4}_{-1.7} )\times 10^{-5}
    \, .     \label{res1}  \eea
where $R=\Gamma(B \rightarrow K^* \gamma)/
\Gamma(B \rightarrow X_s \gamma)$. The formulae used to compute $R$ and 
$BR(B \to K^* \gamma)$ can be found in section \ref{sec:ehn}.
If instead,
 we assume that $T_2$ follows the pole dominance behaviour, we expect a
scaling law of the form $T_1(q^2=0)=T_2(q^2=0) \sim m^{-3/2}$. In this
case we find
\bea T_1(q^2=0)&=& 0.09 \pm 0.01\pm 0.01 \, , \nn \\
R &=& 0.05 \pm 0.02    \, , \nn \\
BR(B \to K^* \gamma)&=& (1.1 \pm 0.3^{+0.4}_{-0.3} )\times 10^{-5}
 \, .    \label{res2}   \eea
The values of $T_1(q^2=0)$ in eqs. (\ref{res1}) and (\ref{res2})
are in good agreement with the results of other similar studies
\cite{hsieh}--\cite{gupta}.
In the first case, eqs. (\ref{res1}),
 the agreement of the lattice predictions with the experimental measurements
is rather satisfactory. In the second case, eqs. (\ref{res2}),
either there is a problem with the
lattice calculations, or the difference may come from long-distance
contributions to the exclusive decay rate \cite{soni}.
Unfortunately, the  present lattice technology cannot estimate these
(and others) long-distance contributions.
Our study of the mass- and  $q^2$-dependence of the form factors favours
 solution (\ref{res1}), even though, given the statistical accuracy of our
results and the systematic errors, we cannot draw any firm conclusion at
this stage.

The systematic difference between  results (\ref{res1}) and (\ref{res2})
 originates from the extrapolation
to large meson masses and small $q^2$, which is needed to obtain the 
physical form factors.
This problem is intrinsic to the lattice approach  at 
 values of the lattice spacing
currently accessible in numerical simulations. In this respect, at present,
the lattice approach is not very different from quark-model calculations.
The $q^2$- and mass-dependence of the form factors, including those relevant to
semileptonic decays, remains a crucial challenge for lattice calculations.

%********************************************************************

\section{Effective Hamiltonian and notation}
\label{sec:ehn}
In this section, we introduce the essential notation
and express the  matrix elements of the effective Hamiltonian and 
the exclusive decay rate in terms of the relevant form factors.
The  operator basis of the effective Hamiltonian density, for weak
radiative $ B$-meson decays, consists of local four-quark operators $O_n$
($n\ =\ 1,2,\dots, 6$) and
magnetic-type operators $O_n$ ($n\ =\ 7,8$) \cite{shif}--\cite{curci2}:
\be{{\cal H} _{{\rm eff}} =-V_{tb}V_{ts}^* G_F/\sqrt{2} \sum_{n=1}^8 
C_n(\mu)O_n(\mu),\label{heff}}\ee
where $C_n$ are the Wilson coefficients.
The  operator that controls the  $b\rightarrow s \gamma$ transition is:
\be{O_7=\Bigl({ e\over 16\pi^2}\Bigr)m_b \left(\bar s
 \sigma^{\mu\nu}b\right)_R F_{\mu\nu}\label{o7}} \, .
\ee
From eqs. (\ref{heff}) and (\ref{o7}),
one finds the matrix element for the transition $B\rightarrow K^*\gamma$
\be{{\cal M}= { e G_F m_b \over  8 \sqrt{2}\pi^2 }
C_7(m_b)V_{tb}V_{ts}^*\epsilon^{\mu *} \langle K^*\vert J_\mu  \vert B \rangle \, , 
\label{M}}\ee
where 
\be{J^\mu \equiv J^\mu_V +J^\mu_A\ =\ \bar s 
\sigma^{\mu\nu}\frac{1\ +\ \gamma_5} {2} q_\nu b \label{jmu}} \ee
and $\epsilon^{\mu}$ and
$q^\mu=p^\mu-k^\mu$ are the photon polarization and momentum 
transfer, respectively.
We parametrize the hadronic matrix element in eq. (\ref{M}) as 
\be{\langle K^*(\eta_r,k)\vert  J^\mu
\vert B(p)\rangle=C_1^\mu T_1(q^2)+iC_2^\mu T_2(q^2)+iC_3^\mu T_3(q^2),
\label{parametrization}}\ee
where
 \bea
C_1^\mu\ &=&\ 2\epsilon^{\mu\alpha\rho\sigma}\eta^*_r(k)_\alpha p_\rho
k_\sigma, \nonumber \\
C_2^\mu\ &=&\ \eta^{\mu \, *}_r(k)
(M_B^2-M_{K^*}^2)-\Bigl(\eta^*_r(k).q\Bigr)(p+k)^\mu,
 \\ 
C_3^\mu\ &=&\ \Bigl(\eta^*_r(k).q\Bigr)\Bigl(\ q^\mu-{q^2\over
M_B^2-M_{K^*}^2}(p+k)^\mu\ \Bigr)\, ; \nonumber \label{ci} \eea 
$T_1$, $T_2$ and $T_3$ are real dimensionless Lorentz-invariant form factors,
and $\eta$ is the $K^*$ polarization vector. 
The vector current $J^\mu_V$  contributes only to $T_1$, 
the axial current  $J^\mu_A$  only to $T_2$
and $T_3$ : 
 \bea
\langle  K^*(\eta_r,k)|J^\mu_V|B(p)
\rangle &=& C_1^\mu T_1(q^2)  \\
\langle  K^*(\eta_r,k)|J^\mu_A|
B(p)\rangle 
&=& iC_2^\mu T_2(q^2)+iC_3^\mu T_3(q^2)\, ; \nonumber \label{separation}\eea
$T_3$ does not contribute to the physical rate, because its coefficient
vanishes for a transversely polarized photon.
By using the relation
$\sigma^{\mu\nu}\gamma^5={i\over 2}\epsilon^{\mu\nu\alpha\beta}\sigma_{\alpha
\beta}$, one finds
\be{ {T_1(0)=T_2(0)} \label{condition}}\ee 
at $q^2=0$. From the matrix element  (\ref{M}), using (\ref{condition}), 
one obtains the  decay width
\be
\Gamma (B\rightarrow K^* \gamma)={\alpha\over 128\pi^4}m_b^2 G_F^2M_B^3
\left(1-{M_{K^*}^
2\over M_B^2}\right)^3|V_{tb}V_{ts}^*|^2C_7(m_b)^2|T_1(0)|^2.\label{decay}
\ee
The physics of this decay is then described by   one form factor only,
$T_1$ at $q^2=0$.

 It is not  convenient  to compare the  theoretical width
(\ref{decay}) with its experimental value, because many  theoretical
uncertainties enter in this quantity: the renormalization scale,
the matching of the lattice to the continuum operators,
the value of $\Lambda_{QCD}$, the possible presence of
new physics beyond the standard model, uncomputed
higher-order corrections,  etc. \cite{ciubsg}.
Now that the inclusive measure is available \cite{cleo2},
it is much more informative to compare instead the
exclusive-to-inclusive
ratio of rates $R$
because in this ratio most of the above-mentioned uncertainties cancel out.
In terms of $T_1$, the ratio $R$ can be written as 
\be R=
\left[ \frac{\Gamma(B \rightarrow K^* \gamma)}
{\Gamma(B \rightarrow X_s \gamma)} \right]^{th}=
\left(\frac{M_B}{m_b}\right)^3\left(1-\frac{M_{K^*}^2}{M_B^2}\right)^3
  \times \frac{4}{1 +
(\lambda_1-9 \lambda_2)/(2 m_b^2)}\times \vert T_1(0) \vert^2 \, ,\label{yyy} \ee
where  we have divided the exclusive rate (\ref{decay})
 by the theoretical inclusive rate,
computed in the HQET parton model.
In  this formalism, the parameters $\lambda_1$ and $\lambda_2$
describe the leading non-perturbative corrections (at order $O(1/m_b^2)$)
to the parton-model predictions for the inclusive rate\footnote{
They are related to the kinetic energy of the {\it b}-quark inside the
{\it B}-meson and to the $B$--$B^*$ mass splitting.} \cite{falk}.

The branching ratio $BR(B \rightarrow K^* \gamma )$ 
is conveniently expressed through a chain of ratios \cite{ciubsg}
\bea
BR(B \rightarrow K^* \gamma ) \, = \, 
R \times
\left[ \frac{\Gamma(B \rightarrow X_s \gamma)}
{\Gamma(B \rightarrow X l \nu_l)} \right]^{th}  \times
 BR^{exp} (B \rightarrow X l \nu_l)
. \label{eq:bre}
\eea
\section{Correlation functions and extraction of the form factors}
\label{subsec:deuxdeux}
In lattice simulations, from the study of two- and three-point correlation functions,
 one extracts the current matrix elements  for a
given momentum transfer and for a given polarization $\eta$ of the $K^*$. 
This is by now a well established technique. 
The reader can  find more details in refs. \cite{abada2} and 
\cite{moredetails}. We follow more closely ref. \cite{abada2}.

In general, the form factors are obtained at values of 
$q^2$ that are constrained by
the quark masses and the values of particle momenta 
accessible on  a given lattice
volume. For this reason, it is necessary to 
extrapolate $T_1$ and $T_2$ to $q^2=0$, in order
to get the physical result. 
In the following,  the different steps of the
procedure to extract the form factors are briefly described\footnote{
We adopt the convention that $B$ and $K^*$ denote
the pseudoscalar heavy meson and the vector light meson, 
whenever no confusion arises.}.

The matrix elements have been computed for an initial meson
at rest and a final vector meson  with momentum $\vec p_{K^*}$.
We have taken $\vec p_{K^*} \equiv  2 \pi / (La)$ 
 $(0,0,0)$, $2 \pi / (La)(1,0,0)$, $2 \pi / (La)(1,1,0)$, 
$2 \pi / (La)(1,1,1)$, and  $2 \pi / (La)(2,0,0)$; where $L$  
is the spatial extension of the lattice, in our case $L\ =\ 18$, and
$a$ is the lattice spacing.
Correlation functions, which are equivalent under the 
hypercubic symmetry, have been averaged.

The initial (final) meson was created (annihilated)
by  using the  pseudoscalar ($J_B=\bar b \gamma_5 q$) and
  local vector ($J^\alpha_{K^*}=\bar q \gamma^\alpha s$) densities,
 inserted at times $t_B/a=28$
and $t_{K^*}=0$, respectively.  We have varied the time position of $J_\mu$ in the
interval $t_J/a=10$--$14$.

In order to obtain the hadronic matrix element, 
the  following procedure has been used:
\begin{enumerate}
\item We have computed masses and source
matrix elements by fitting the two-point  correlation functions
at large time distances to the expressions given below.
 For the pseudoscalar meson, we have used:
\bea
C_B(t_B)\ \equiv \ 
\langle  0\vert J_B(t_B) J^\dagger_B(\vec x=0, t=0)\vert 0\rangle 
= \frac{Z_B}{2 M_B} e^{- M_B t_B},
\label{c55m}
\eea 
where $J_B(t_B)=
\int d^3 x  J_B(\vec x , t_B)$ and $Z_B^{1/2}= \langle 0
 \vert J_B \vert B(\vec p_B=\vec 0)\rangle$.
For the vector current, one has 
\bea
C_{K^*}^{\alpha\beta}(\vec p_{K^*} , t_{K^*})& \ =\ &
\langle 0 \vert J_{K^*}^\alpha( \vec p_{K^*}, t_{K^*})
J_{K^*}^{\dagger \beta}(\vec x =0, t=0) \vert 0\rangle  \nonumber\\ 
& \ =\ & 
 \Bigl( g^{\alpha \beta}-\frac{ p_{K^*}^{\alpha} \, p_{K^*}^{\beta} } { M_{K^*}^2 } \Bigr)
\frac{Z_{K^*}}  { 2 E_{K^*} } e^{-E_{K^*} \, t_{K^*}} \, ,\label{cvm}
\eea
where \bea J_{K^*}^\alpha(\vec p_{K^*},t_{K^*})&=&
\int d^3 x \ e^{-i\vec p_{K^*}\cdot {\vec x}} J_{K^*}^\alpha (\vec x , t_{K^*}) \, .  \eea
 In eq. (\ref{cvm}), $Z_{K^*}$ is defined through
\be
\eta^{\beta *}_r\sqrt{Z_{K^*}}=
\langle K^*(\eta_r ,\vec  p_{K^*}) \vert J_{K^*}^\beta \vert0\rangle 
\, .\label{zv}\ee
\item We have then extracted the matrix elements from the ratios
\bea 
R_{\mu\alpha}=\frac{C_{3}^{\mu\alpha}(t_{B},t_J,\vec p_{K^*})}
{C_{K^*}^{{\alpha\alpha}}(\vec p_{K^*},t_{J})
C_{B}(t_B-t_{J})}\, , \label{rapport}
\eea
with
\bea &\,&
C_{3}^{\mu\alpha}(t_{B},t_{J},\vec p_{K^*})=
\langle 0 \vert J_{K^*}^\alpha(\vec p_{K^*},0)
 J^\mu(-\vec p_{K^*},t_J)
J_B^\dagger(t_B)
\vert 0 \rangle = \nonumber \\ 
&\,&
\sum_{ r}^{}\eta^{\alpha}_r{\sqrt{Z_{K^*}}\over
2E_{K^*}}e^{- E_{K^*}t_{J}}{\sqrt{Z_B}\over 2M_B}e^{-
M_{B}(t_{B}-t_J)}\nonumber \\ 
&\times& 
\langle K^*(\eta_r,\vec p_{K^*}) \vert J^\mu
\vert B\rangle \, ,\label{c3e}
\eea
\end{enumerate}
where the magnetic operator $J^\mu$, renormalized in the continuum,  is given in terms of the corresponding lattice bare
operator by $J^\mu = Z_\sigma 
J^\mu_{\rm latt}$, with $Z_\sigma=0.98$ taken from
perturbation theory \cite{carlotta}.
To obtain the matrix elements, we have used two different
procedures, denoted by ``analytic'' and ``ratio'' methods, which  
 are discussed in detail in ref. \cite{abada2}.  In the
``ratio'' method,  for each fixed-time distance,  the
three-point correlation function is divided by the two-point functions of
the $B$ and $K^{*}$ mesons (with corresponding momentum) as in eq. (\ref{rapport}), 
in order to cancel the exponential time-dependence. 
The ``analytic'' method differs from the
previous one in that, instead of dividing by the numerical two-point correlation
functions,  we divide by the
corresponding analytic expressions  using  the source matrix elements 
($Z_B$ and $Z_{K^*}$) and the
meson energies   obtained  from the
fit of the two-point functions  at zero momentum.
The energy of the $K^*$ si computed from  the meson mass,
as explained  in  eqs. (\ref{eq:enl})
and (\ref{eq:enc}) below.
When both the initial and final mesons are at 
rest, and at large time distances, the two methods are
practically equivalent and lead to almost identical results. However, when
the meson momenta are different from zero
the two methods are expected to agree only up to $O(a)$ effects. 
\par  In refs.  \cite{gupta,moredetails,nieves},
 it was found that discretization
errors appear to be reduced, if one uses, for
the two-point correlation
functions,   the  ``lattice" dispersion relation of a free boson 
\begin{equation} \bar C(t,\vec p) = \frac{Z}{2 \sinh E} e^{-E t},
\label{eq:emr} \end{equation}
where 
\begin{equation} E=\frac{2}{a} {\rm arc sinh}
\left( \sqrt{{\rm sinh}^2 \left( \frac{m a}{2}\right)
 +\sum_{i=1,3}{\rm sin}^2\left(\frac
{p_i a}{2}\right)}
\right) \label{eq:enl} \end{equation}
 and $\vec p$ is the momentum of the meson. The same is true in our
case. We have verified this point by studying  
 the ratio  ${\cal R}(t)= C(t,\vec p)/\bar C(t,\vec p )$, where  
 $C(t, \vec p)$ is the two-point correlation function of a meson with
momentum $\vec p$, as computed in our  simulation, and   $\bar C(t, \vec p)$
is given by the expression in eq. (\ref{eq:emr}), with $Z$ and $m$ taken 
from the fit of  the zero-momentum correlation to $\bar C(t, \vec p=\vec 0)$.
At large time distances, we find that ${\cal R}(t)$ is   closer to 1
(typically $1.05 \pm 0.01$ instead of $1.10 \pm 0.02$),
if we use eq. (\ref{eq:emr}) instead of the standard expression
\begin{equation} \hat C(t,\vec p) = \frac{Z}{2 E} e^{-E t},\label{eq:hatC}
 \end{equation}
with
\begin{equation} E=\sqrt{m^2 + \vert \vec p\vert ^2}.\label{eq:enc}
\end{equation} 
For this reason, throughout our analysis, we have  fitted the two-point
 functions to $\bar C(t,\vec p )$, eq. (\ref{eq:emr}),  in addition to the ``standard"
form $\hat C(t,\vec p)$,  eq. (\ref{eq:hatC}).  
 \par 
Using $\bar C(t, \vec p)$, the ``ratio" and ``analytic"
methods yield only slightly different results (see section \ref{sec:setup}).

\section{$q^2$-dependence of the form factors and scaling laws}
\label{sec:q2d}

In order to obtain the form factors at the physical point,
we need to extra\-po\-la\-te  
in both mass and momentum. The final results  depend critically
on the assumptions on the $q^2$- and heavy mass-dependence, which deserve
a detailed discussion.
\par At fixed $\vec p_{K^*}$, with  $\vert \vec p_{K^*}\vert\ll  M_B$ in the
$B$-meson  rest frame,  the following scaling laws can be derived \cite{wise}:
\begin{equation}
\label{scala} 
\frac{T_1}{\sqrt{M_B}}=\gamma _{1}\times \left( 1+\frac{\delta _{1}}{M_B}
+ \dots \right)\, ,
 \qquad  T_2\sqrt{M_B}=\gamma _2\times \left( 1+\frac{\delta _2}{M_B}
+ \dots \right) \, ;
\end{equation}
these are valid up to logarithmic corrections. 
As mentioned in the introduction, ``scaling"
laws
for the form factors at $q^2=0$ can only be found by using extra assumptions
on the $q^2$-dependence of the form factors.  Such a  procedure is acceptable,
provided the ``scaling" laws derived in this way  respect the
 \underline{exact} condition
$T_1(0)=T_2(0)$. This is a non-trivial constraint: 
 the $q^2$-behaviour of $T_1$ and $T_2$
has to compensate the different mass dependence of the two form factors 
near the zero recoil point, see eq. (\ref{scala}). Thus, for example, the
popular assumption of pole dominance for both $T_1$ and $T_2$  would give
 $T_1(0) \sim M_B^{-1/2}$ and $T_2(0) \sim M_B^{-3/2}$, which is inconsistent
with eq. (\ref{condition}).\par 
In  all lattice simulations, which try to compute 
$B$-meson form factors by extrapolating from lower heavy-quark masses, 
we are forced to
make assumptions on the corresponding $q^2$-dependence. This is a consequence
of the fact that  the extrapolation in
the mass, at fixed light-meson momentum,  pushes the values of $q^2$ towards $q^2_{max}$.
 This problem can only be avoided by going to much  smaller values of the lattice spacing,
thus allowing a large increase in the range of accessible masses and momenta.
The assumptions on  the $q^2$-dependence of the form factors can be
tested, although in a small domain of 
 momenta, directly on the numerical results.
\begin{figure}[t]
\vspace{9pt}
\begin{center}\setlength{\unitlength}{1mm}
\begin{picture}(160,100)(-5,25)
%\put(0,-55){\special{postscript}}
$$\epsfbox{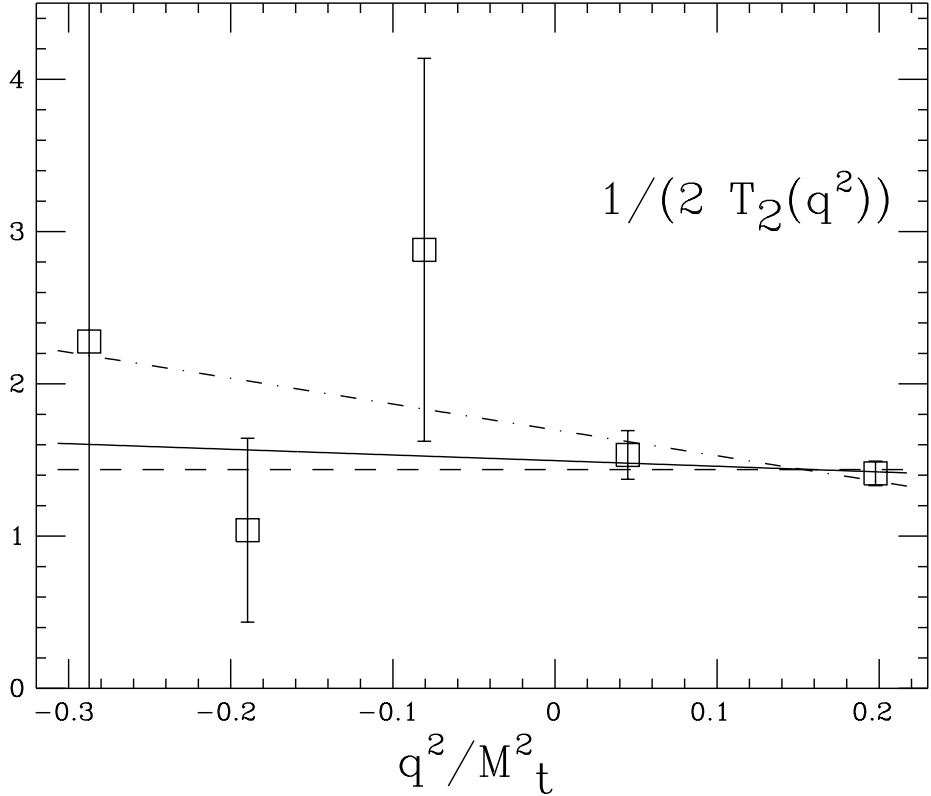}$$
\end{picture}
\end{center}
\caption{\it{The inverse form factor $1/(2T_2(q^2))$ is shown as a function
of $q^2/M_t^2$ for $K_H=0.1200$. The form factor has  already been
extrapolated in the mass of the light quark to the strange-quark mass.
 The dash-dotted  line
represents the pole-dominance behaviour,
with the mass of the axial meson $M_t$ taken from the fit of the
two-point correlation function; the solid line corresponds to a 
 pole-dominance behaviour, with $M_t$ left as a free parameter; the
dashed line is a fit with $T_2(q^2)$ constant in $q^2$.}}
\label{fig:t2}
\end{figure}

In fig.  \ref{fig:t2}, we show  $1/(2 T_2(q^2))$ as
a function of the dimensionless variable $q^2/M_t^2$, where $M_t$ is the 
appropriate mass for the axial 
meson exchanged in the $t$-channel, in the pole-dominance approximation. 
If pole dominance is valid, the  form factor should have a linear behaviour
in $q^2$, with a slope proportional to $1/M_t^2$. 
In the case of $T_2$, we found that
the mass extracted from the slope is much larger 
(by a factor of order 2 in most of the cases)
than the mass that we have obtained 
from the axial two-point correlation function. 
In other words, $T_2(q^2)$ is  flatter than predicted by pole dominance.
In the case of $T_1$, the point at momentum $(0,0,0)$ cannot be computed.
The absence of this point and the large statistical and systematic
errors   at large  momenta make it difficult to test
the validity of the pole-dominance hypothesis  directly on our
results. A consistency check will be provided
is section \ref{sec:setup}.
Although our data
are not accurate enough  to draw a definite conclusion, 
they suggest that  assuming  $T_2$ almost constant in $q^2$ and $T_1$ following
pole dominance  gives a good description of our results. 
This assumption is consistent
with the ``scaling" laws described above. We call this option $m^{-1/2}$-scaling.
The $m^{-1/2}$-scaling is similar to what has been encountered in the 
QCD sum-rules calculation \cite{bbd,ball} of the semileptonic $B \to \rho$ form factors
$V$ and $A_1$, which are related to $T_1$ and $T_2$ in the large-mass
limit.
In fig.  \ref{fig:t2}, we also give the curves corresponding to 
a pole dependence for $T_2$ ($m^{-3/2}$-scaling).
This possibility is also consistent with the ``scaling" laws, if a dipole 
dominance behaviour is assumed for $T_1$, but it is
disfavoured by our numerical results, as discussed above and shown in the
figure. For these data,  the  uncorrelated $\chi^2$, corresponding to the
fit of $T_2$ to a constant, is $\chi^2= 0.65$ to be compared to
$\chi^2= 1.15$, in the case of the fit of $T_2$ to a pole dominance behaviour.
Similar results have been obtained at the other values of the heavy
quark mass.
It is clear, that  an infinite number of possible $q^2$-dependence for
$T_{1,2}$, which are compatible with the scaling laws, can be found. 
However it would be
impossible to distinguish among them, given the statistical errors, the
systematic uncertainty in the extraction of the form factors, the
effects of $O(a)$,  and the limited range in $q^2$ and $M_B$.
Thus we take the two possibilities,  $m^{-1/2}$-scaling and $m^{-3/2}$-scaling,
as representative of a full class of ``scaling" laws. In section
 \ref{sec:setup},
we  will see that the two options lead to  quite  different results for
the physical value of $T_1(0)$, which enters in the calculation of the
$B \to K^* \gamma$ rate.
\par There 
remains to discuss the dependence on the light-quark masses (spectator
and active). 
 At fixed heavy-quark mass and light-meson momentum  $\vec p_{K^*}$,
the generic form factor $F$ ($F\ =\ T_1$, $T_2$) has been
extrapolated linearly in the mass of the active light quark
\be F \ =\ \alpha + \beta m_q \, , \label{lq} \ee
to values corresponding to the physical strange meson  $K^*$,
assuming  the form factors  independent of the mass of the
spectator quark. This is in agreement with the results
of  ref. \cite{ukqcd}, where it was shown that  
the dependence of the form factors on the mass 
 of the spectator quark is very small.
We believe that the extrapolation in the light-quark mass is  unlikely to
be a source of an important uncertainty
within the present statistical accuracy.

\section{Lattice setup and numerical results}
\label{sec:setup}
\begin{table}
\centering
\begin{tabular} {|c|c|c|c|c|c|}
\hline
Form factor& $K_H$& {\rm r-sinh }  & {\rm a-sinh } & 
{\rm r-stand}   & {\rm a-stand }   \\
\hline 
$T_1$ & 0.1330  & 0.35(3) & 0.34(3) & 0.33(3) & 0.34(3) \\
$T_2$ & 0.1330  & 0.37(2) & 0.37(2) & 0.35(2) & 0.35(2) \\
$\bar T_2$ & 0.1330  & 0.37(2) & 0.36(2) & 0.35(2) & 0.35(2) \\
\hline
$T_1$ & 0.1250  & 0.34(4) & 0.33(3) & 0.31(3) & 0.32(3) \\
$T_2$ & 0.1250  & 0.36(2) & 0.36(2) & 0.33(2) & 0.33(2) \\
$\bar T_2$ & 0.1250  & 0.34(3) & 0.33(3) & 0.31(3) & 0.32(3) \\
\hline
$T_1$& 0.1200  & 0.33(4) & 0.32(3) & 0.30(3) & 0.31(3) \\
$T_2$ &0.1200  & 0.35(2) & 0.35(2) & 0.32(2) & 0.32(2) \\
$\bar T_2$ &0.1200  & 0.33(5) & 0.32(4) & 0.29(4) & 0.31(4) \\
\hline
$T_1$ & 0.1150  & 0.33(4) & 0.32(4) & 0.29(3) & 0.30(3) \\
$T_2$ & 0.1150  & 0.35(2) & 0.35(2) & 0.30(2) & 0.30(2) \\
$\bar T_2$ & 0.1150  & 0.32(6) & 0.31(6) & 0.28(5) & 0.30(6) \\
\hline
$T_1$ & $B$  & 0.24(3) & 0.23(3) & 0.20(3) & 0.21(3) \\
$T_2$ & $B$  & 0.25(2) & 0.25(2) & 0.22(2) & 0.22(2) \\
$\bar T_2$ & $B$  & 0.22(5) & 0.21(5) & 0.19(5) & 0.21(5) \\
\hline
\end{tabular}
\caption{\it{$T_1$, $T_2$  and $\bar T_2$ at $q^2=0$ for different values
of the heavy-quark masses, as extracted using the pole and constant
behaviour respectively.  We give in
 the second column the hopping parameter of the heavy quark
($B$ denotes the extrapolation to the $B$-meson
using the $m^{-1/2}$ scaling law);  the
 light-quark mass has been extrapolated to the strange quark mass;
the values of the form factors obtained with the ratio method from
the sinh or standard fits (r-sinh and r-stand) or with the analytic method
from the sinh or standard fits (a-sinh and a-stand) are shown from
the third to the sixth columns.}}
\label{tab:TunTdeux}
\end{table}
\begin{table}
\centering
\begin{tabular} {|c|c|c|c|c|c|}
\hline
Form factor& $K_H$& {\rm r-sinh }  & {\rm a-sinh } & 
{\rm r-stand}   & {\rm a-stand }   \\
\hline 
$T_1$ & 0.1330  & 0.42(4) & 0.40(4) & 0.39(4) & 0.40(4) \\
$T_2$ & 0.1330  & 0.36(2) & 0.36(2) & 0.34(2) & 0.34(2) \\
\hline
$T_1$ & 0.1250  & 0.35(4) & 0.34(3) & 0.32(3) & 0.33(3) \\
$T_2$ & 0.1250  & 0.32(2) & 0.31(2) & 0.29(2) & 0.29(2) \\
\hline
$T_1$& 0.1200  & 0.31(4) & 0.31(3) & 0.28(3) & 0.29(3) \\
$T_2$ &0.1200  & 0.29(2) & 0.29(2) & 0.26(2) & 0.27(2) \\
\hline
$T_1$ & 0.1150  & 0.28(4) & 0.28(3) & 0.25(3) & 0.26(3) \\
$T_2$ & 0.1150  & 0.27(2) & 0.27(2) & 0.24(2) & 0.24(2) \\
\hline
$T_1$ & $B$  & 0.10(1) & 0.09(1) & 0.08(1) & 0.09(1) \\
$T_2$ & $B$  & 0.09(1) & 0.09(1) & 0.08(1) & 0.08(1) \\
\hline
\end{tabular}
\caption{\it{$T_1$ and  $T_2$   at $q^2=0$ for different values
of the heavy-quark masses, as extracted using the dipole and pole
behaviour respectively. 
We give in
 the second column the hopping parameter of the heavy quark
($B$ denotes the extrapolation to the $B$-meson using the
$m^{-3/2}$ scaling law);  the
 light quark mass has been extrapolated to the strange-quark mass;
the values of the form factors obtained with the ratio method from
the sinh or standard fits (r-sinh and r-stand) or with the analytic method
from the sinh or standard fits (a-sinh and a-stand) are shown from 
the third to the sixth column.}}
\label{tab:dipole}
\end{table}
\begin{table}
\centering
\begin{tabular} {|c|c|c|c|c|}
\hline
\multicolumn{5}{|c|}{$T(0)=T_1(q^2=0)=T_2(q^2=0)$}\\
\hline
 Scaling law& {\rm r-sinh }  & {\rm a-sinh } & 
{\rm r-stand}   & {\rm a-stand }   \\
\hline 
$m^{-1/2}$   & 0.25(2) & 0.25(2) & 0.21(2) & 0.22(2) \\
$m^{-3/2}$   & 0.09(1) & 0.09(1) & 0.08(1) & 0.08(1) \\
\hline
\end{tabular}
\caption{\it{$T(0)=T_1(q^2=0)=T_2(q^2=0)$  extrapolated to the
$B$-meson using the $m^{-1/2}$ and the $m^{-3/2}$
scaling laws; 
the values of the form factors obtained with the ratio method from
the sinh or standard fits (r-sinh and r-stand) or with the analytic method
from the sinh or standard fits (a-sinh and a-stand) are shown from 
the second to the fifth columns.}}
\label{tab:combined}
\end{table}

The numerical simulation was performed on the 
6.4 gigaflops APE machine, at $\beta \ = \ 6.0$, on an
 $18^3\times 64$ lattice, using the 
SW-Clover action \cite{clover}\ in the quenched approximation.
The results were obtained from a sample of 170 gauge configurations.
The statistical errors have been estimated by a jacknife method, by decimating
10 configurations from the total set.
For each configuration we have computed the quark propagators for 7 different
values of the Wilson hopping parameter $K_W$, corresponding to
\lq\lq heavy" quarks, $K_H=0.1150$, $0.1200$, $0.1250$, $0.1330$,
and \lq\lq light" quarks, $K_L=0.1425$, $0.1432$ and $0.1440$.
In order to extract masses and source matrix elements, we have fitted the
heavy and light meson two-point functions to eqs. (\ref{c55m})
and (\ref{cvm})  in the time interval $14 \le t/a \le 28$ and
$10 \le t/a \le 28$ respectively. The relevant matrix elements of $O_7$ have
been extracted, for different polarizations and momenta
of the $K^*$, from  the three-point function, see eq. (\ref{c3e}),
 with the operator inserted  at $10 \le t_J/a \le 14$.
We found that   the
critical value of $K_L$, corresponding to the chiral limit, is
$ K_{cr}=0.14545(1)$; the inverse lattice spacing, obtained from 
$m_\rho$, is  $a^{-1}=1.96(7)$ GeV; the value of the Wilson parameter
for the  strange quark, determined from the mass of the kaon,  is
 $ K_s=0.1435(1)$.
In the following, the labels ``ratio" and ``analytic"  refer 
respectively  to the ratio and analytic method as explained in section 
\ref{subsec:deuxdeux}. The numbers quoted with ``sinh" and ``standard"
are obtained  respectively with  the fits 
to eqs. (\ref{eq:emr}) and  (\ref{eq:hatC}) 
for the two-point functions (see section \ref{subsec:deuxdeux}).
\par
We now explain the analysis of the scaling behaviour of the form factors
that, on the basis of the above discussion,
has been done in combination with the study of their $q^2$-dependence.
Unless otherwise stated, the form factors are those obtained  after the
 extrapolation in the active light quark mass to the value of
 the hopping parameter corresponding to the strange quark.

\begin{enumerate} \item
According to HQET, when the mass of the meson is sufficiently large,
$T_2$ at zero recoil  should follow the behaviour given in eq. (\ref{scala}).
A fit of  our data for $T_2(q^2_{max})$ to 
$M^{\alpha}_B ( a + b/M_B)$, with $\alpha$, $a$ and $b$ as free parameters,
gives $\alpha = -0.41\,(10)$,
 and $b/a = -350\,(70)$ MeV ($\alpha = -0.68\,(10)$,
 and $b/a = -480\,(50)$ MeV) with the sinh-fit (standard-fit).
The  exponent $\alpha$ is thus compatible with the value of $-1/2$
  predicted by HQET, see eq. (\ref{scala}). To reduce the number
of parameters, we have then extrapolated 
$T_2(q^2_{max})$ in $1/M_B$, with 
 $\alpha=-1/2$, using  eq. (\ref{scala}). In this case, we obtain
$\gamma_2=20.2\, (1.6)$ MeV$^{1/2}$ and  $\delta_2=-430\, (50)$ MeV
($\gamma_2=16.8\, (1.4)$ MeV$^{1/2}$ and  $\delta_2=-320\, (50)$ MeV). 
These parameters
 correspond to the following values of  $T_2(q^2_{max})$ for the $B$-meson
\bea
 & \hbox{sinh} &\,\,\,\,\,\,\,\, \hbox{standard} \nn \\ 
T_2(q^2_{max},B) = & 0.25(2) &\,\,\,\,\,\,\,\, 0.22(2)\, .\\ \nn
\eea
This result is
 also reported  in table \ref{tab:TunTdeux} as $T_2$ with the label $B$.
\item Let us assume the ``pure-$m^{-1/2}$" behaviour for the form factors.
By  ``pure-$m^{-1/2}$" behaviour, we mean that $T_2(q^2)$ is independent  of
$q^2$ (this is compatible with our results, cf. fig. 
\ref{fig:t2})
and that $T_1(q^2)$ follows a pole-dominance behaviour, 
with the mass of the vector
meson as measured from the two-point correlation functions. In table
\ref{tab:TunTdeux}, we  
compare, for all the values of  the heavy-quark mass,
$T_2=T_2(q^2=0)=T_2(q^2_{max})$ with $
T_1=T_1(q^2=0)=T_1(q^2) \times(1-q^2/M^2_t)$. The values of $T_1$ and $T_2$
are compatible within the statistical errors and the differences coming from
 different extrapolations (analytic, ratio, sinh, standard).
\item In order to check the stability of the results, we also performed a 
linear extrapolation in $q^2$ of $T_2(q^2)$ to $q^2=0$, 
at fixed heavy-quark mass, followed by  a fit of $T_2(q^2=0)$ to 
eq. (\ref{scala}). The values obtained in this way are reported as
$\bar T_2$ in table \ref{tab:TunTdeux}. Since we have not done any assumption
on the $M_B$-dependence of the slope of the fit, this procedure is not
a priori incompatible with the scaling laws discussed in the previous section.
 It should be noticed that,
since we have an extra parameter in the fits, i.e.  the linear slope in
$q^2$, the extrapolated value $\bar T_2$ has a larger error.  
\item Even though our results prefer a flat $q^2$-dependence for $T_2$,
we have also fitted $T_1$ and $T_2$ by
assuming a pure  ``$m^{-3/2}$" behaviour 
for the form factors. This means that we first fit the $q^2$-dependence 
of the form factors
to $T_1(q^2)=T_1/(1-q^2/M^{\prime\,  2}_t)^2$ and
$T_2(q^2)=T_2/(1-q^2/M^{2}_t)$, where
for each value of the heavy-quark
mass, the value of $M_t^\prime$ and $M_t$
 are those  computed from the corresponding two-point functions. The form 
factors extracted in this way are reported in tab.  \ref{tab:dipole}.
We have then extrapolated in $1/M_B$ using the expressions 
$T_1 \, M_B^{3/2}= \chi_1 \times (1 + \theta_1/M_B)$ and 
$T_2 \, M_B^{3/2}= \chi_2 \times (1 + \theta_2/M_B)$.
\item We notice that, for the values of the heavy quark masses 
at which  we have
computed the form factors, the value of
$T_1(q^2=0)$ ($T_2(q^2=0)$) 
extracted by assuming the pole (constant)
$q^2$-behaviour  differs at most by $20$--$25\%$ from the value
obtained by assuming the dipole- (pole-) $q^2$-depencence,
as can be read  in tables 
\ref{tab:TunTdeux} and \ref{tab:dipole}.
The same would be true  if we  extrapolated the form factors using
the two different scaling laws to the charm-quark mass.  
The values extrapolated
to the $B$-meson instead, differ by about a factor of 2. 
\item
One should also take into account the distortion in the mass dependence
of the form factors  due to lattice artefacts at large quark masses.
This effect has been measured non-perturbatively both by the APE 
collaboration 
at $\beta=6.0$ \cite{tassosb}  and the UKQCD collaboration
at $\beta=6.2$  \cite{nievesb}.
Only with more accurate data, and exploring a much larger range of masses and
momenta at larger values of $\beta$, 
will it be possible to decide the fundamental issue of the scaling behaviour.
\item We can exploit the information that the two form factors must be equal
at $q^2=0$, by making a combined fit of $T_1$ and $T_2$,
with the constraint $T(0)=T_1(q^2=0)=T_2(q^2=0)$. This can be done both in the 
$m^{-1/2}$ and $m^{-3/2}$ cases.  
The results are given in table \ref{tab:combined}.
\item
As a consistency check of our results, we have also inverted the order of the fits, by
fitting first $T_2(q^2_{\rm max})$ in $1/M_B$ and then assuming the $q^2$ behaviour of
the form factor. The results are, 
within the errors, well compatible with those reported 
in tables \ref{tab:TunTdeux} and \ref{tab:dipole}.
\end{enumerate}
From the results reported in tables \ref{tab:TunTdeux}--\ref{tab:combined},
we quote
\bea T_1(q^2=0) &=& 0.23(2)(2)       \qquad \hbox{scaling } m^{-1/2}\, , \nn \\
T_1(q^2=0) &=&  0.09(1)(1)  \qquad \hbox{scaling } m^{-3/2}\, , \eea
from which we have derived the results of  (\ref{res1}) and (\ref{res2})
in the introduction.

\section{Conclusion}
We have computed the form factor relevant for $B \rightarrow K^* \gamma$ decays. In order
to extract the physical form factor from the lattice results, we have extrapolated in
momentum transfer and in the mass of the heavy quark.  The results, corresponding to different
choices of the scaling law, can differ by about a factor of 2. 
Within large uncertainties,
our study suggests that the scaling law ``$m^{-1/2}$", by which $T_2(q^2)$ has a very small
dependence on the momentum transfer, is preferred, in agreement with the results of ref. \cite{gupta}.
We cannot exclude, however, the ``$m^{-3/2}$" scaling behaviour, or any intermediate solution.
In order to improve the accuracy of the predictions,
it is necessary to be able to work with heavier quark masses and to increase the
range of $q^2$. This can only be achieved  by going to larger lattices.

\section*{Acknowledgements}    
We thank M. Ciuchini, J. Flynn, R. Gupta,  A. Le Yaouanc, J. Nieves, O. P\`ene
and A. Soni,
for interesting discussions on the subject of this paper.
  G.M. acknowledges partial support from 
M.U.R.S.T.  We also acknowledge partial support by the EC contract
CHRX-CT92-0051.


\begin{thebibliography}{999}
\bibitem{cleo1} R. Ammar et al. (CLEO collaboration), 
Phys. Rev. Lett. 71 (1993) 674.
\bibitem{cleo2} B. Barish et al. (CLEO collaboration),
 invited talk at the ICHEP94, Glasgow, Scotland, July 1994,
CLEO-CONF-94-1, to appear in the Proceedings; CLNS-94-1314, Dec 1994.
\bibitem{hqet} A. Ali, T. Mannel, and T. Ohl, Phys. Lett. B298 (1993)  195.
\bibitem{Dominguez} C.A. Dominguez et al., Phys. Lett. B214 (1988) 459. 
\bibitem{colangelo} P. Colangelo et al., Phys. Lett. B317 (1993) 183.
\bibitem{bball} P. Ball, TUM-T31-43-93, (hep-ph 9308244).
\bibitem{narison} S. Narison, Phys. Lett. B327 (1994) 354.  
\bibitem{ali} A. Ali et al., CERN-TH.7118/93, 
MPI-Ph/93-97, DESY 93-193 (hep-ph 9401277).
\bibitem{qm1} N.G. Deshpande et al., Z. Phys. C40  (1988) 369.
\bibitem{qm2} P.J. O'Donnel and H.K.K. Tung, Phys. Rev. D44  (1991) 741.
\bibitem{qm3} T. Altomari, Phys. Rev. D37 (1988)  677.
\bibitem{hsieh} C.W. Bernard et al., Nucl. Phys. (Proc Suppl) B26
 (1992) 347; Phys. Rev. Lett. 72 (1994) 1402.
\bibitem{ukqcd} K.C. Bowler et al.  (UKQCD collaboration), Phys. Rev.
Lett. 72 (1994)
 1398; SHEP 93/94-29;  B. Gough,  talk given at LATTICE94, 
Bielefeld, Germany, October 1994 
 (hep-lat9411086), to appear in the Proceedings;
D.R. Burford et al. (UKQCD collaboration), SHEP 95-09 (hep-lat9503002).
\bibitem{gupta} T. Bhattacharya and R. Gupta, talk
given  at LATTICE94, Bielefeld, Germany, October 1994, 
(hep-lat9501016), to appear in the Proceedings.
\bibitem{abadaglasgow} As. Abada, invited talk at the ICHEP94,
 Glasgow, Scotland, July 1994,
LPTHE Orsay-94/79 (hep-9409338), to appear in the Proceedings; Ph. Boucaud,
talk given at LATTICE94, Bielefeld,
 Germany, October 1994, LPTHE Orsay-94/109 (hep-lat9501015),
to appear in the Proceedings.  
\bibitem{wise} N. Isgur and M.B. Wise, Phys. Rev. D42 (1990) 2388.
\bibitem{donoghue} G. Burdman and J.F. Donoghue, Phys. Lett. B270 (1991) 55.
\bibitem{alain} R. Aleksan, A. Le Yaouanc, L. Oliver, O. P\`ene and 
J.C. Raynal,  DAPNIA/SPP/94-24, LPTHE-Orsay 94/15. 
\bibitem{autres} A. Ali, V.M. Braun and H. Simma, Z. Phys. C63 (1994) 437.
\bibitem{colangelo2} P. Colangelo,  F. De Fazio
 and P. Santorelli, BARI-TH/94-174
, DSF-T-94/12, INFN-NA-UIV-94/12.
\bibitem{ciubsg} M. Ciuchini, E. Franco, G. Martinelli, L. Reina and L.
Silvestrini, Phys. Lett. B334 (1994) 137. 
\bibitem{soni} D.  Atwood, B.  Blok and A.  Soni,
SLAC-PUB-6635, Aug. 1994 (hep-ph-9408373).
\bibitem{golo} E. Golowich and S. Pakvasa,
UMHEP-411, Aug. 1994 (hep-ph-9408370).
\bibitem{taiw} H.Y.  Cheng, IP-ASTP-23-94 (hep-ph/9411330). 
\bibitem{lubicz} V. Lubicz, G. Martinelli and C.T. Sachrajda,
Nucl. Phys. B356 (1991) 301.
\bibitem{shif} M.A. Shifman, A.I. Vainshtein and V.I. Zakharov,
Phys. Rev.  18 (1978) 2583.
\bibitem{grinstein} B. Grinstein, R. Springer and B. Wise, 
Phys. Lett. B202 (1988) 138; Nucl. Phys. B339  (1990) 269.
\bibitem{altri}
R. Grigjanis, P.J. O'Donnel, M. Sutherland and H. Navelet, Phys. Lett.
 B213 (1988) 355;  B223 (1989) 239; B237 (1990) 252;
G. Cella, G. Curci, G. Ricciardi and A. Vicer\'e, Phys. Lett.  B248
(1990) 181;
M.  Misiak, Phys. Lett. B269 (1991) 161, B321 (1994) 193;
Nucl. Phys. B393 (1993) 23.
\bibitem{silve1} M. Ciuchini, E. Franco, G. Martinelli, L. Reina and L.
Silvestrini, Phys. Lett.  B316 (1993) 127.
\bibitem{silve2} M. Ciuchini, E. Franco, L. Reina and L. Silvestrini,
Nucl. Phys. B421 (1994) 41.
\bibitem{curci2} G. Cella, G. Curci, G. Ricciardi and A. Vicer\'e,
Phys. Lett.  B325 (1994) 227.
\bibitem{falk} A.F. Falk, M. Luke and M.J. Savage,
Phys. Rev. D49 (1994) 3367.
\bibitem{abada2}   As. Abada et al., Nucl. Phys. B416 (1994) 675.
\bibitem{moredetails}  C.R. Allton et al. (APE collaboration), 
CERN-TH.7484/94 (hep-lat/9411011),  to appear in Phys. Lett. B.
\bibitem{carlotta} A. Borrelli, C. Pittori, R. Frezzotti and E. Gabrielli,
Nucl. Phys. B409 (1993) 382.
\bibitem{nieves} L. Lellouch  et al. (UKQCD collaboration),
EDINBURGH-94-548 (hep-lat/9410013), to be published in Nucl. Phys. B.
\bibitem{bbd}  P. Ball, V.M. Braun and H.G. Dosch,  Phys. Rev. D44 (1991) 3567.
\bibitem{ball}  P. Ball, Phys. Rev. D48 (1993) 3190.
\bibitem{clover}B. Sheikholeslami and R. Wohlert, Nucl. Phys. B259 (1985) 572. 
\bibitem{tassosb}  A. Vladikas (APE collaboration),
  talk given at LATTICE94,  Bielefeld, Germany, October 1994, 
 (hep-lat/9502012), to appear in the Proceedings. 
\bibitem{nievesb} J. Nieves (UKQCD collaboration),
 talk given at LATTICE94,  Bielefeld, Germany, October 1994, 
 (hep-lat/9412013), to appear in the Proceedings. 
\end{thebibliography}
\end{document}